\documentstyle[twoside,fleqn,espcrc2,epsfig]{article}
\newcommand{\link}[2]{  U_{ #1 , #2}  }
\newcommand{\linkdagger}[2]{  U_{ #1 , #2}^\dagger  }
\newcommand{\kronecker}[2]{  \delta_{#1 \; #2}  }

\newcommand{\apart}[3]{      
\sum_{#3} \Big( \link{#1}{#3} \kronecker{#1}{#2-\hat{#3}} 
                -\linkdagger{#1-\hat{#3}}{#3} 
                                        \kronecker{#1}{#2+\hat{#3}} \Big) }

\newcommand{\bparta}[3]{
\sum_{#3} \Big( \link{#1}{#3} \link{#1+\hat{#3}}{#3} \link{#1+2 \hat{#3}}{#3}
\kronecker{#1}{#2- 3 \hat{#3}}\\
&&~~- \linkdagger{#1-\hat{#3}}{#3} \linkdagger{#1-2\hat{#3}}{#3}
\linkdagger{#1-3 \hat{#3}}{#3} \kronecker{#1}{#2+ 3 \hat{#3}} \Big) }

\newcommand{\bpartb}[4]{
\sum_{#3} \sum_{#4\ne#3}\bigg[ \Big( \link{#1}{#3} \link{#1+\hat{#3}}{#4} \link{#1+\hat{#3} + \hat{#4}}{#4} \kronecker{#1}{#2-\hat{#3}-2\hat{#4}} \\
&&~~~~~~ - \linkdagger{#1-\hat{#4}}{#4} \linkdagger{#1-2\hat{#4}}{#4} 
     \linkdagger{#1- \hat{#3} - 2\hat{#4}}{#3} \kronecker{#1}{#2+\hat{#3}+2\hat{#4}}\Big)\\
&&~~~+ \Big( \link{#1}{#4} \link{#1+\hat{#4}}{#4} \link{#1+
2\hat{#4}}{#3} \kronecker{#1}{#2-\hat{#3}-2\hat{#4}} \\
&&~~~~~~ - \linkdagger{#1-\hat{#3}}{#3} \linkdagger{#1-\hat{#3}-\hat{#4}}{#4} 
     \linkdagger{#1- \hat{#3} - 2\hat{#4}}{#4}\kronecker{#1}{#2+\hat{#3}+2\hat{#4}}\Big)\\
&&~~~+ \Big( \linkdagger{#1-\hat{#4}}{#4} \linkdagger{#1-2\hat{#4}}{#4}
\link{#1-2\hat{#4}}{#3} \kronecker{#1}{#2-\hat{#3}+2\hat{#4}} \\
&&~~~~~~ - \linkdagger{#1-\hat{#3}}{#3} \link{#1-\hat{#3}}{#4} \link{#1- \hat{#3} +
\hat{#4}}{#4} \kronecker{#1}{#2+\hat{#3}-2\hat{#4}}\Big)\\
&&~~~+ \Big( \link{#1}{#3} \linkdagger{#1+\hat{#3}-\hat{#4}}{#4}
\linkdagger{#1+\hat{#3}-2\hat{#4}}{#3} \kronecker{#1}{#2-\hat{#3}+2\hat{#4}}\\
&&~~~~~~ - \link{#1}{#4} \link{#1+\hat{#4}}{#4} \linkdagger{#1- \hat{#3} +
2\hat{#4}}{#3} \kronecker{#1}{#2+\hat{#3}-2\hat{#4}}           
          \Big) \bigg] }

\renewcommand{\arraystretch}{2}

\newcommand{\be}{\begin{eqnarray*}}
\newcommand{\ee}{\end{eqnarray*}}
\newcommand{\bc}{\begin{center}}
\newcommand{\ec}{\end{center}}
\newcommand{\bfg}{\begin{figure}[h]}
\newcommand{\efg}{\end{figure}}
\newcommand{\nn}{\nonumber}
\newcommand{\AmS}{{\protect\the\textfont2
  A\kern-.1667em\lower.5ex\hbox{M}\kern-.125emS}}

\title{Staggered Fermion Actions with Improved Rotational Invariance
    \thanks{ We thank
       the Deutsche Forschungsgemeinschaft for the support of the work 
    under grants Pe 340/3-3, Pe 340/6-1}}
\author{A. Peikert with B. Beinlich, A. Bicker, F. Karsch 
and E. Laermann \address{Fakult\"at f\"ur Physik, Universit\"at Bielefeld, 33501 Bielefeld, Germany}}

\begin{document}

\begin{abstract}
We introduce a class of improved actions for staggered fermions 
which to ${\cal O}(p^4)$ and ${\cal O}(p^6)$, respectively, lead to 
rotationally invariant propagators. We discuss the resulting reduction of 
flavour symmetry breaking in the meson spectrum and comment on 
the improvement in the calculation of thermodynamic observables.
\end{abstract}

\maketitle

\section{Introduction}
In finite temperature calculations of QCD the cut-off dependence in the high
temperature phase can be reduced with the ${\cal O}(a^2)$ improved Naik 
action. The energy density of an ideal fermi gas only deviates by $20 \%$ from
the continuum value on
  lattices with temporal extent $N_{\tau}=4$ to be compared with a deviation of
  $70 \%$ for the standard staggered action\cite{engels}. But still a further
improvement is necessary to achieve an accuracy of a few percent as for the 
pure gauge calculations.\\
Another lattice artifact is the breaking of flavour
symmetry at ${\cal O}(a^2)$ which is not improved in the 
Naik formulation\cite{bernard}. An improvement by a factor of $2$ can be
achieved by introducing fat links in the standard fermion action\cite{blum}.\\[-4mm]
\section{Construction of fermion actions with improved rotational invariance}  
A general ansatz for a free fermion action with a difference scheme
of arbitrary high order is given by
\be
S_{F}&=&\sum_{x, \nu > \mu}~ \eta_{x,\mu}~\bar{\psi}(x) ~\sum_{j>0,k,l,m}
c_{j,k,l,m} \nn\\
&&\cdot \Big[~\psi(x+ja_{\mu}+ka_{\nu}+la_{\rho}+ma_{\sigma}) \nn\\
&&-\psi(x-ja_{\mu}-ka_{\nu}-la_{\rho}-ma_{\sigma}) \Big] ,\\
\ee\\[-6mm]
where $j$ is odd and $k,l,m$ are even to respect staggered symmetries. 
$\eta_{x,\mu}=(-1)^{(x_0+..+x_{\mu-1})}$ and $\eta_{x,0}=
  1$ denote the staggered phases.\\
The coefficients can be fixed by demanding that the fermion propagator is
rotationally invariant up to ${\cal O}(p^4)$ or ${\cal O}(p^6)$ respectively.\\[2mm]
$\bullet$ ${\cal O}(p^4)$ improved propagator\\[1mm]
For actions including 1-link and 3-link paths in
  the difference scheme we obtain the constraints:\\[-4mm]
\be
c_{1,0,0,0} + ~3~c_{3,0,0,0} + 6 ~c_{1,2,0,0}&=&{1 / 2}\nn\\
c_{1,0,0,0} + 27~c_{3,0,0,0} + 6 ~c_{1,2,0,0}&=&24~c_{1,2,0,0}\nn\\
\ee\\[-8mm]
The fermion action {$ S_{F}= \bar{\Psi}  M \Psi $} then has the form  
\begin{eqnarray*}
M[U]_{ij}&=& m~\delta_{ij} + \eta_{i}\cdot \Big(c_{1,0,0,0} ~{ A[U]_{ij}} \\
&& + c_{3,0,0,0} ~{ B_1[U]_{ij}}+  c_{1,2,0,0} ~{ B_2[U]_{ij}} \Big)\\[2mm]
{A[U]_{ij}}
&=&\apart{i}{j}{\mu}\\
{B_1[U]_{ij}} 
&=&\bparta{i}{j}{\mu}\\[-4mm]
\end{eqnarray*}
\begin{eqnarray*}
\lefteqn{B_2[U]_{ij} =}\\
&&\bpartb{i}{j}{\mu}{\nu}\\
\end{eqnarray*}
Choosing $c_{1,2,0,0}=0$ yields the Naik action:\\[-5mm]
\be
c_{1,0,0,0}= {9 / 16}~~~~~c_{3,0,0,0}= -{1 / 48} \nn\\
\ee\\[-7mm]
Choosing the linear 3-link term to be zero we obtain the so-called p4 action:\\[-5mm]
\be
c_{1,0,0,0}= {3 / 8}~~~~~c_{1,2,0,0}= {1 / 48} \nn\\[-2mm]
\ee
$\bullet$ ${\cal O}(p^6)$ improved propagator:\\
For actions including up to 7-link paths of euclidean length up to $\sqrt{13}$
in the difference scheme we find four equations which constrain the six
coefficients.\\
In Table 1 we show two possible choices of the $c_{i,j,k,l}$ , the first one,
denoted by p6m, corresponds to the minimal set of four
non-vanishing parameters, the second one, called p6, contains two
free parameters which can be used for tuning. The coefficients of the fixed
point action (fp) which are shown in the third column are similar to those of
the rotational invariant actions i.e. the value of the coefficients is to a
large extent determined by rotational invariance.\\[3mm]
\renewcommand{\arraystretch}{0.95}
 \begin{tabular}{|l|l|l|c|}\hline
    $(i,j,k,l)$&\multicolumn{3}{|c|}{$c_{i,j,k,l}$} \\
    \hline
    ~&\multicolumn{1}{c|}{p6m} &\multicolumn{1}{c|}{p6} &\multicolumn{1}{c|}{fp \cite{biet}} \\
    \hline
    \hline
    (1,0,0,0)  & 0.3375    & 0.32      &  0.3308695 \\
    (1,2,0,0)  & 0.01875   & 0.02      &  0.0220591 \\
    (1,2,2,0)  & 0.0023438 & 0.0010938 &  0.0023284 \\
    (3,0,0,0)  & 0.0072917 & 0.0047917 &  0.0117443 \\
    (1,2,2,2)  &           & 0.00125   &  0.0002419 \\
    (3,2,0,0)  &           & 0.00125   & -0.0002466 \\
    (....)     &           &           &  ...       \\ 
    \hline
  \end{tabular}
  {\small Table 1. Coefficients for ${\cal O}(p^6)$ rotational invariant actions and 
    fixed point action.}

\section{Thermodynamic properties and zero temperature dispersion relations}
In pure gauge theory improvement of the high temperature behavior with
tree level improved actions leads to a large improvement even close to $T_c$.
It is expected that the same will hold for full QCD when an improved fermion
action with an improved high temperature limit is used.\\
In Figure 1 we show the free fermion energy density for various actions and note that the
p4 action has deviations from the continuum value of maximum $8\%$ (for $N_{\tau}=6$).\\
\epsfig{file=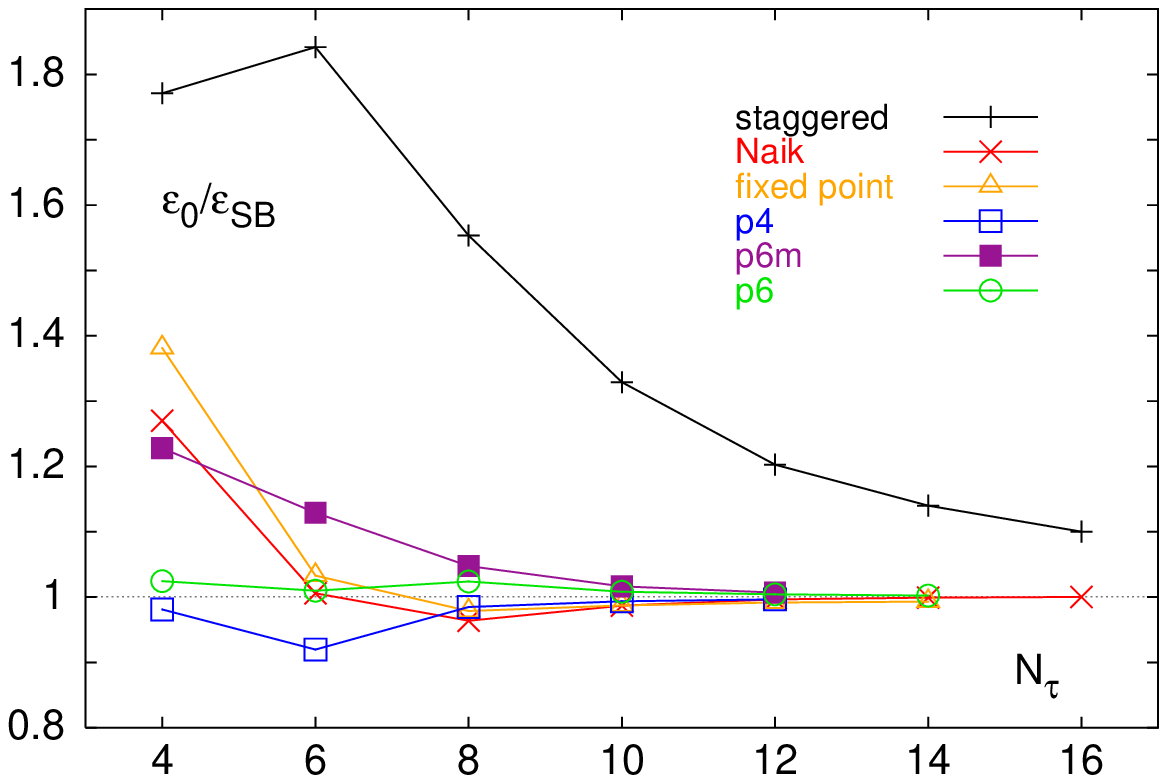,width=78mm}
{\vspace*{-7mm}{~\newline}}
{\small Figure 1. Deviations from the ideal gas value for various fermion actions.}\\[2mm]
The dispersion relation, $E=E(p_x,p_y,p_z)$, results from the poles of the
propagator, $D^{-1}(iE,\vec{p})=0$. The p4 action dispersion relation shown in Figure 2
is close to the continuum form  $E(p)=p$ for a wide range of momenta.\\  
\epsfig{file=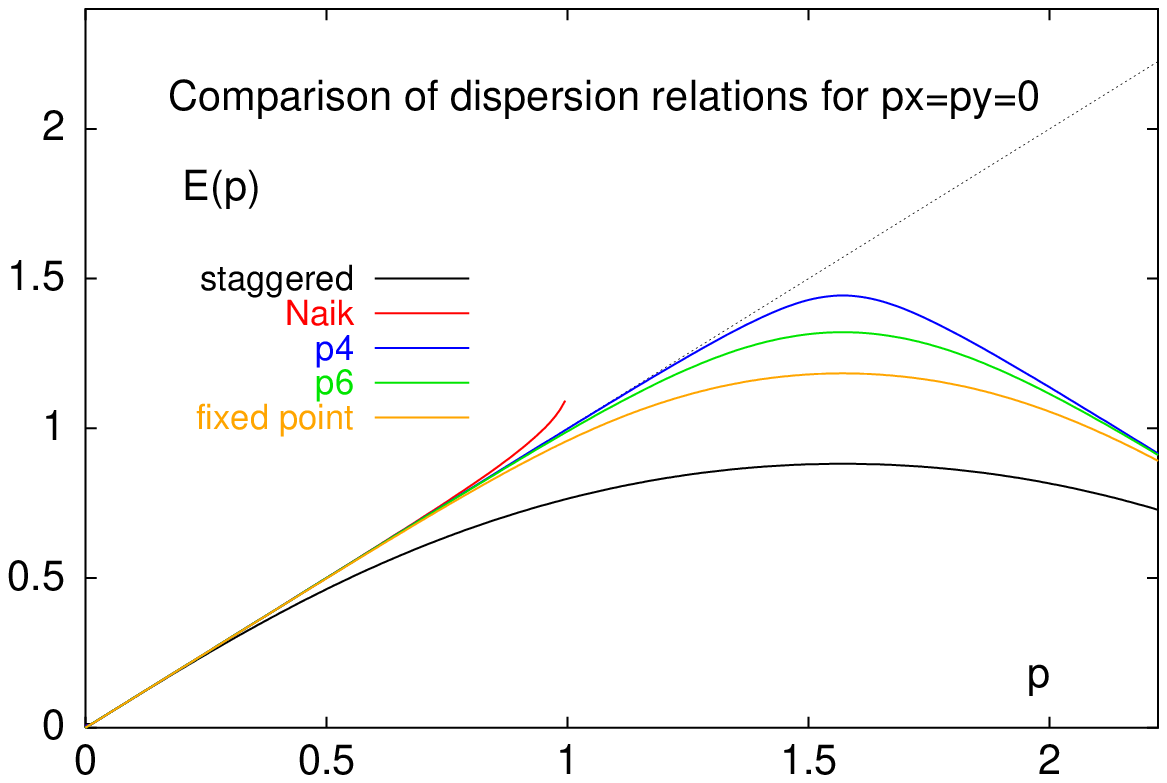,width=78mm}
{\vspace*{-7mm}{~\newline}}
{ \small Figure 2. Zero temperature dispersion relations for several actions.}\\[-5mm]
\renewcommand{\arraystretch}{1.08}
\begin{figure*}[t]
  \begin{minipage}{12.2cm}
    \begin{tabular}{|l|l|l|l|l|l|l|}\hline
      action& \multicolumn{1}{c|}{$\omega$} & \multicolumn{1}{c|}{$m_{\pi}$} &
      \multicolumn{1}{c|} {$m_{\pi_{2}}$} & \multicolumn{1}{c|}{$m_{\rho}$} & 
      \multicolumn{1}{c|}{$m_{\pi}/m_{\rho}$} & $\Delta_{\pi}$\\
      \hline
      staggered          & 0.0 & 0.598(1) & 0.792(7) & 0.979(15) & 0.613(10) & 0.199(8)\\
      p4                 & 0.0 & 0.643(1) & 0.838(6) & 1.015(15) & 0.633(10) & 0.192(7)\\
      p6                 & 0.0 & 0.676(1) & 0.870(9) & 1.032(15) & 0.655(10) & 0.188(9)\\
      staggered          & 0.2 & 0.571(1) & 0.645(3) & 0.888(15) & 0.643(11) & 0.086(4)\\
      staggered          & 0.4 & 0.569(1) & 0.635(3) & 0.865(15) & 0.658(12) & 0.076(4)\\
      p4$_{\mbox{fat1}}$ & 0.2 & 0.619(1) & 0.694(3) & 0.915(20) & 0.676(15) & 0.082(4)\\
      p4$_{\mbox{fat2}}$ & 0.2 & 0.609(1) & 0.693(4) & 0.905(20) & 0.673(15) & 0.093(5)\\
      \hline
    \end{tabular}
  \end{minipage}
  \begin{minipage}{3.2cm}
    {\small Table 2. \\ Meson masses, $\pi/\rho$ mass ratio and pion splitting for
    various actions at a bare quark mass of $m_q a = 0.05$.}\vfill
  \end{minipage}
\end{figure*}
\renewcommand{\arraystretch}{1.0}

\section{Testing flavour symmetry - Numerical results}
We have performed a quenched simulation using a tree level improved 1$\times$2 
action at $\beta=4.10$ on a $16^3\times30$ lattice. We calculated the meson propagators from wall
sources on 57 configurations for several fermion actions including also fat links \cite{blum}.\\
An indicator of flavour symmetry breaking is the mass splitting of the Goldstone
 pion $\pi$ and the non-Goldstone pion $\pi_2$
normalized by the rho meson mass, $\Delta_{\pi}=\left(m_{\pi_2}-m_{\pi}\right) / m_{\rho}$. \\
\epsfig{file=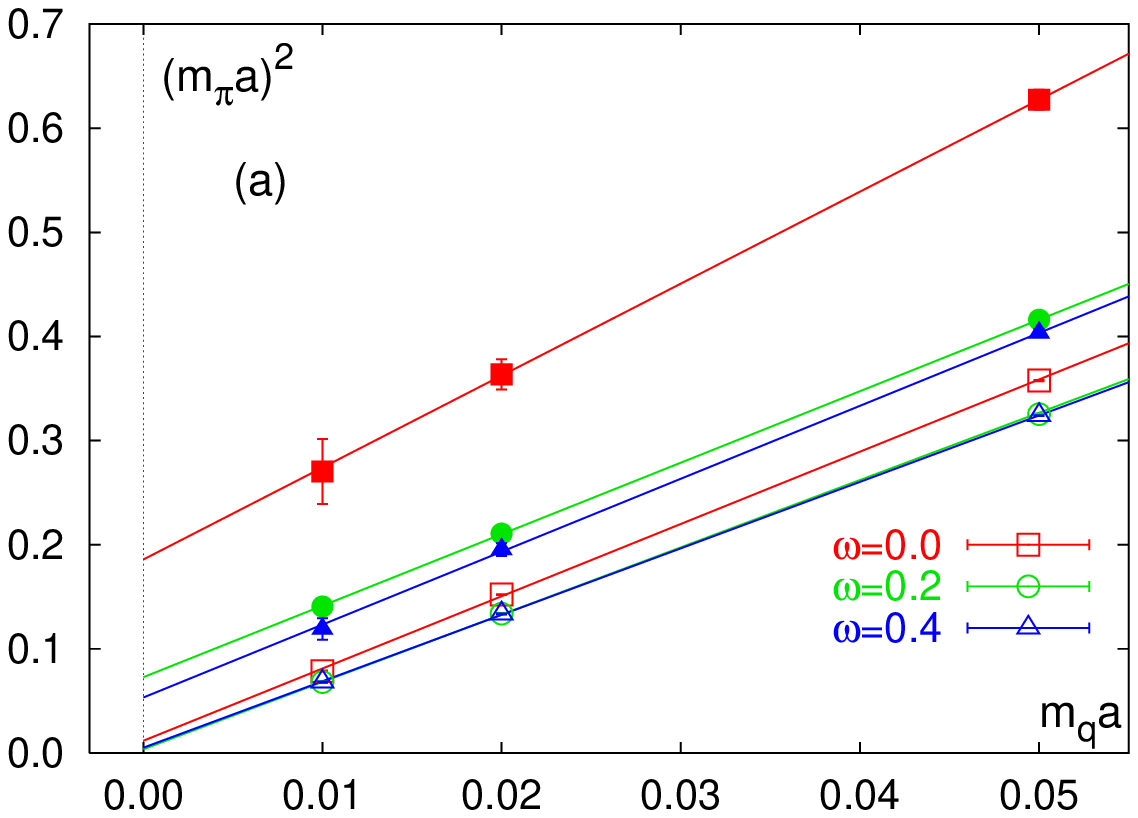,width=78mm}
\epsfig{file=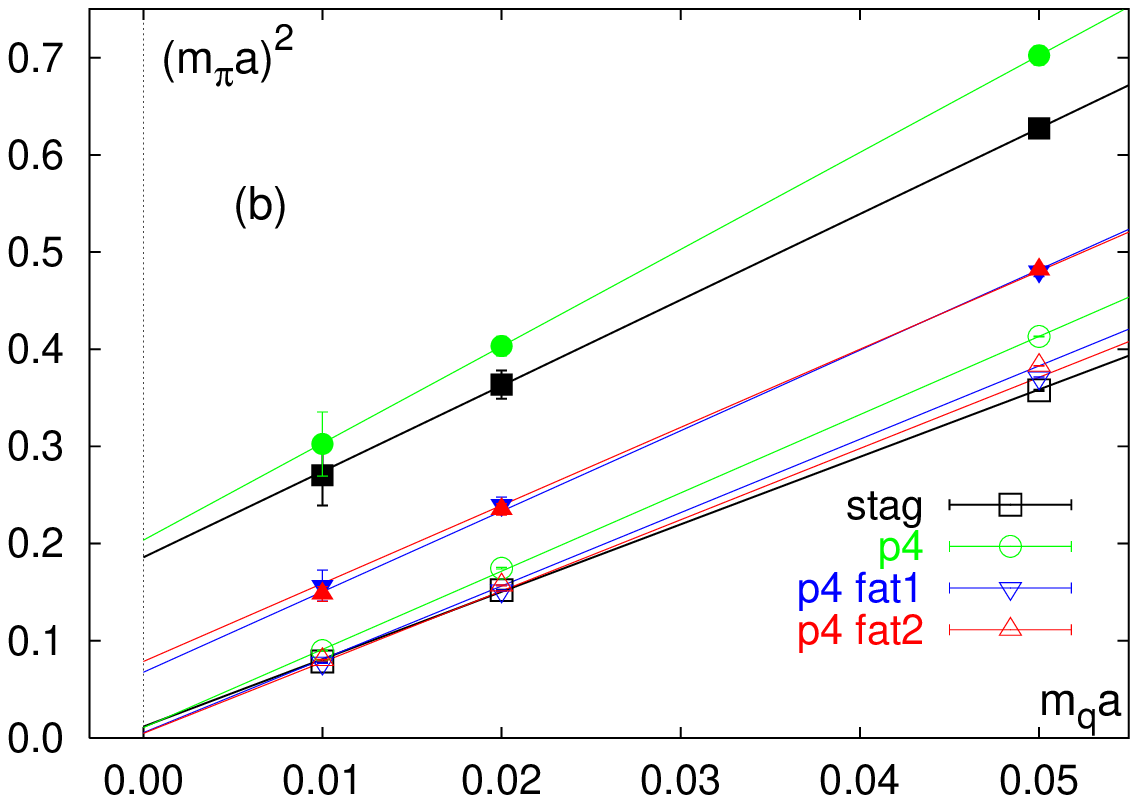,width=78mm}
{\small Figure 3. The squared pion mass vs the bare quark mass (a) for standard fat
and non-fat actions and (b) for fat p4 and non-fat standard and p4 actions.
Unfilled symbols are the Goldstone, filled symbols the non-Goldstone pions.}\\[1mm]
In Table 2 we show results of the pion splitting for the following actions with a bare quark mass
$m_q a = 0.05$ :\\[-6mm]
\begin{itemize}
\itemsep -4mm
\item standard staggered, p4 and p6 action\\
\item standard staggered action with fat links for $\omega =0.2$ and $\omega =0.4$\\
\item p4 action with fat links in the 1-link path (p4$_{\mbox{fat1}}$) and 
p4 action with fat links in the 1-link and 3-link path (p4$_{\mbox{fat2}}$) for $\omega = 0.2$\\[-5mm]
\end{itemize}
One finds that for the p4 and p6 actions the pion splitting is
not reduced significantly, whereas fat p4 actions lead to an improvement by a
factor of 2 similar to what has been observed for the standard action.
In Figure 3 we show the squared pion mass vs the bare quark mass with a linear
extrapolation to zero quark mass. We do not see any significant improvement of flavour
symmetry for the tree level improved gauge action compared to the Wilson gauge 
action. For the fat p4 action the improvement of flavour symmetry persists in
the chiral limit.\\[-6mm]
\section{Conclusions}
The p4 action using fat links in the
one-link term is a good candidate for finite temperature
simulations in full QCD. Both thermodynamic properties and flavour symmetry
are improved significantly. \\[-5mm]


\begin{thebibliography}{9}

\bibitem{engels}
J. Engels {\it et al.}, Phys.Lett. B396 (1997) 210
\bibitem{bernard}
C. Bernard {\it et al.} (MILC coll.), Nucl.Phys. B
    (Proc. Suppl.) 53 (1997) 212
\bibitem{blum}
T. Blum {\it et al.}, Phys.Rev. D55 (1997) 1133 
\bibitem{biet}
W. Bietenholz et al. , Nucl.Phys.B (Proc. Suppl.) 53 (1997) 921
\end{thebibliography}
\end{document}